\documentstyle[12pt]{article}

 1

\catcode`\@=11 
\def\lsim{\mathrel{\mathpalette\@versim<}}
\def\gsim{\mathrel{\mathpalette\@versim>}}
\def\@versim#1#2{\vcenter{\offinterlineskip
        \ialign{$\m@th#1\hfil##\hfil$\crcr#2\crcr\sim\crcr } }}

\catcode`@=11
\def\chkspace{%
  \relax   
  \begingroup\ifhmode\aftergroup\dochksp@ce\fi\endgroup}
\def\dochksp@ce{%
  \unskip              
  \futurelet\chkspct@k\d@chkspc  
}
\def\d@chkspc{%
  \let\nxtsp@ce=\relax
  \ifx\chkspct@k.\else     
    \ifx\chkspct@k,\else
      \ifx\chkspct@k;\else
        \ifx\chkspct@k!\else
          \ifx\chkspct@k?\else
            \ifx\chkspct@k:\else
              \ifx\chkspct@k)\else
              \ifx\chkspct@k(\else
                \ifx\chkspct@k]\else
                  \ifx\chkspct@k-\else
                    \ifx\chkspct@k\egroup\else  
                      \let\nxtsp@ce=\put@space  
                    \fi
                  \fi
                \fi
              \fi
              \fi
            \fi
          \fi
        \fi
      \fi
    \fi
  \fi
  \nxtsp@ce
}
\def\put@space{$\;$}
\catcode`@=12
 
\def\ra{{$\rightarrow$}\chkspace}

\def\adhoc{{\it ad hoc}\chkspace}
\def\ie{{\it i.e.}\chkspace}

\def\ep{{e$^+$e$^-$}\chkspace}

\def\gluino{\relax\ifmmode \tilde{g} \else $\tilde{g}$ \fi\chkspace}
 
\def\qq{q$\overline{\rm q}$\chkspace}

\def\pp{p$\overline{\rm p}$\chkspace}
 
\def\bb{\relax\ifmmode {\rm b}\bar{\rm b}
       \else ${\rm b}\bar{\rm b}$ \fi\chkspace}
\def\cc{\relax\ifmmode {\rm c}\bar{\rm c}
       \else ${\rm c}\bar{\rm c}$ \fi\chkspace}
\def\tt{\relax\ifmmode {\rm t}\bar{\rm t}
       \else ${\rm t}\bar{\rm t}$ \fi\chkspace}

\def\qqg{\relax\ifmmode {\rm q}\overline{\rm q}{\rm g}
\else q$\overline{\rm q}$g \fi\chkspace}

\def\afb{\relax\ifmmode A_{FB} \else
{{$A_{FB}$}}\fi\chkspace}
\def\afbb{\relax\ifmmode A_{FB}^b \else
{{$A_{FB}^b$}}\fi\chkspace}
\def\pafb{\relax\ifmmode \tilde{A}_{FB} \else
{{$\tilde{A}_{FB}$}}\fi\chkspace}
\def\pafbb{\relax\ifmmode \tilde{A}_{FB}^b \else
{{$\tilde{A}_{FB}^b$}}\fi\chkspace}
 
\def\pafbzo{\relax\ifmmode \tilde{A}_{FB}|_{O(0)} \else
{{$\tilde{A}_{FB}|_{O(0)}$}}\fi\chkspace}
\def\pafbfo{\relax\ifmmode \tilde{A}_{FB}|_{\oalp} \else
{{$\tilde{A}_{FB}|_{\oalp}$}}\fi\chkspace}
\def\pafbso{\relax\ifmmode \tilde{A}_{FB}|_{\oalpsq} \else
{{$\tilde{A}_{FB}|_{\oalpsq}$}}\fi\chkspace}
\def\pafbto{\relax\ifmmode \tilde{A}_{FB}|_{\oalpc} \else
{{$\tilde{A}_{FB}|_{\oalpc}$}}\fi\chkspace}
 
\def\pafbbzo{\relax\ifmmode \tilde{A}_{FB}^b|_{O(0)} \else
{{$\tilde{A}_{FB}^b|_{O(0)}$}}\fi\chkspace}
\def\pafbbfo{\relax\ifmmode \tilde{A}_{FB}^b|_{\oalp} \else
{{$\tilde{A}_{FB}^b|_{\oalp}$}}\fi\chkspace}
\def\pafbbso{\relax\ifmmode \tilde{A}_{FB}^b|_{\oalpsq} \else
{{$\tilde{A}_{FB}^b|_{\oalpsq}$}}\fi\chkspace}
\def\pafbbto{\relax\ifmmode \tilde{A}_{FB}^b|_{\oalpc} \else
{{$\tilde{A}_{FB}^b|_{\oalpc}$}}\fi\chkspace}
 
\def\afbo0{\tilde{A}_{FB}|_{O(0)}}
\def\afbo1{\tilde{A}_{FB}|_{\oalp}}
\def\afbo2{\tilde{A}_{FB}|_{\oalpsq}}
\def\afbo3{\tilde{A}_{FB}|_{\oalpc}}
 
\def\lam{\relax\ifmmode \Lambda_{\overline{MS}}
       \else {{$\Lambda_{\overline{MS}}$}}\fi\chkspace}
\def\lamuds{\relax\ifmmode \Lambda^{(3)}_{\overline{MS}}
       \else {{$\Lambda^{(3)}_{\overline{MS}}$}}\fi\chkspace}
\def\lamudsc{\relax\ifmmode \Lambda^{(4)}_{\overline{MS}}
       \else $\Lambda^{(4)}_{\overline{MS}}$\fi\chkspace}
\def\lamudscb{\relax\ifmmode \Lambda^{(5)}_{\overline{MS}}
       \else $\Lambda^{(5)}_{\overline{MS}}$\fi\chkspace}

\def\alp{\relax\ifmmode \alpha_s\else $\alpha_s$\fi\chkspace}
\def\alpbar{\relax\ifmmode \overline{\alpha_s}
       \else $\overline{\alpha_s}$\fi\chkspace}
\def\alpmz{\relax\ifmmode \alpha_s(M_Z)\else $\alpha_s(M_Z)$\fi\chkspace}
\def\alpmzsq{\relax\ifmmode \alpha_s(M_Z^2)
       \else $\alpha_s(M_Z^2)$\fi\chkspace}
 
\def\oalp{\relax\ifmmode O(\alpha_s)\else{{O($\alpha_s$)}}\fi\chkspace}
\def\oalpsq{\relax\ifmmode O(\alpha_s^2)
           \else{{O($\alpha_s^2$)}}\fi\chkspace}
\def\oalpc{\relax\ifmmode O(\alpha_s^3)
           \else{{O($\alpha_s^3$)}}\fi\chkspace}

\def\npb{Nucl. Phys.\chkspace}

\def\prd{Phys. Rev.\chkspace}

\def\z0{{$Z^0$}\chkspace}
\def\Dst{\relax\ifmmode {\rm D}^* \else {D$^*$}\fi\chkspace}
\def\Dpl{\relax\ifmmode {\rm D}^+ \else {D$^+$}\fi\chkspace}
\def\D0{\relax\ifmmode {\rm D}^0 \else {D$^0$}\fi\chkspace}
\def\Kst{\relax\ifmmode {\rm K}^* \else {K$^*$}\fi\chkspace}
\def\K0{\relax\ifmmode {\rm K}^0_s \else {K$^0_s$}\fi\chkspace}
\def\Kpl{\relax\ifmmode {\rm K}^+ \else {K$^+$}\fi\chkspace}
\def\Kstz{\relax\ifmmode {\rm K}^{*0} \else {K$^{*0}$}\fi\chkspace}

\makeatletter
\def\@seccntformat#1{\csname the#1\endcsname.\hskip 1em}

\makeatother
 
\begin{document}

\newcommand{\y}{{\it y \/}}
\newcommand{\Dymax}{$\Delta\y_{max}$~}
\newcommand{\Dycent}{$\Delta\y_{cent}$~}
\newcommand{\Dy}{$<\Delta\y>$~}
\newcommand{\yzero}{$\y_{o}$~}
\newcommand{\absy}{$|\y|$~}
 
\newcommand{\epem}{e$^+$e$^-$~}
\newcommand{\tautau}{$\tau^+$$\tau^-$~}
 

\thispagestyle{empty}
 
\centerline{\bf FIRST STUDY OF RAPIDITY GAPS IN \epem ANNIHILATION}
 
\vspace {1.0cm}
 
%
%
%
  \def\iADEL{$^{(1)}$}
  \def\iBOL{$^{(2)}$}
  \def\iBU{$^{(3)}$}
  \def\iBRUN{$^{(4)}$}
  \def\iCIT{$^{(5)}$}
  \def\iUCSB{$^{(6)}$}
  \def\iUCSC{$^{(7)}$}
  \def\iCIN{$^{(8)}$}
  \def\iCSU{$^{(9)}$}
  \def\iCOLO{$^{(10)}$}
  \def\iCOL{$^{(11)}$}
  \def\iFER{$^{(12)}$}
  \def\iFRA{$^{(13)}$}
  \def\iILL{$^{(14)}$}
  \def\iLBL{$^{(15)}$}
  \def\iMIT{$^{(16)}$}
  \def\iMASS{$^{(17)}$}
  \def\iMISS{$^{(18)}$}
  \def\iMOSC{$^{(19)}$}
  \def\iNAG{$^{(20)}$}
  \def\iOREG{$^{(21)}$}
  \def\iPAD{$^{(22)}$}
  \def\iPERU{$^{(23)}$}
  \def\iPISA{$^{(24)}$}
  \def\iRUT{$^{(25)}$}
  \def\iRAL{$^{(26)}$}
  \def\iSOGANG{$^{(27)}$}
  \def\iSLAC{$^{(28)}$}
  \def\iTENN{$^{(29)}$}
  \def\iTOH{$^{(30)}$}
  \def\iVAND{$^{(31)}$}
  \def\iWASH{$^{(32)}$}
  \def\iWISC{$^{(33)}$}
  \def\iYALE{$^{(34)}$}
  \def\dead{$^{\dag}$}
  \def\andgen{$^{(a)}$}
  \def\andper{$^{(b)}$}
%
%
\begin{center}
\mbox{K. Abe                 \unskip,\iNAG}
\mbox{K. Abe                 \unskip,\iTOH}
\mbox{I. Abt                 \unskip,\iILL}
\mbox{T. Akagi               \unskip,\iSLAC}
\mbox{N.J. Allen             \unskip,\iBRUN}
\mbox{W.W. Ash               \unskip,\iSLAC$^\dagger$}
\mbox{D. Aston               \unskip,\iSLAC}
\mbox{K.G. Baird             \unskip,\iRUT}
\mbox{C. Baltay              \unskip,\iYALE}
\mbox{H.R. Band              \unskip,\iWISC
% ...
}
\it
  \vskip \baselineskip                   
  \centerline{(The SLD Collaboration)}   
  \vskip \baselineskip                   
%
%
%
  \iADEL
     Adelphi University,
     Garden City, New York 11530 \break
  \iBOL
     INFN Sezione di Bologna,
     I-40126 Bologna, Italy \break
  \iBU
     Boston University,
     Boston, Massachusetts 02215 \break
  \iBRUN
     Brunel University,
     Uxbridge, Middlesex UB8 3PH, United Kingdom \break
  \iCIT
     California Institute of Technology,
     Pasadena, California 91125 \break
  \iUCSB
     University of California at Santa Barbara,
     Santa Barbara, California 93106 \break
  \iUCSC
     University of California at Santa Cruz,
     Santa Cruz, California 95064 \break
  \iCIN
     University of Cincinnati,
     Cincinnati, Ohio 45221 \break
  \iCSU
     Colorado State University,
     Fort Collins, Colorado 80523 \break
  \iCOLO
     University of Colorado,
     Boulder, Colorado 80309 \break
  \iCOL
     Columbia University,
     New York, New York 10027 \break
  \iFER
     INFN Sezione di Ferrara and Universit\`a di Ferrara,
     I-44100 Ferrara, Italy \break
  \iFRA
     INFN  Lab. Nazionali di Frascati,
     I-00044 Frascati, Italy \break
  \iILL
     University of Illinois,
     Urbana, Illinois 61801 \break
  \iLBL
     Lawrence Berkeley Laboratory, University of California,
     Berkeley, California 94720 \break
  \iMIT
     Massachusetts Institute of Technology,
     Cambridge, Massachusetts 02139 \break
  \iMASS
     University of Massachusetts,
     Amherst, Massachusetts 01003 \break
  \iMISS
     University of Mississippi,
     University, Mississippi  38677 \break
  \iMOSC
     Moscow State University,
     Institute of Nuclear Physics,
     119899 Moscow,
     Russia    \break
  \iNAG
     Nagoya University,
     Chikusa-ku, Nagoya 464 Japan  \break
  \iOREG
     University of Oregon,
     Eugene, Oregon 97403 \break
  \iPAD
     INFN Sezione di Padova and Universit\`a di Padova,
     I-35100 Padova, Italy \break
  \iPERU
     INFN Sezione di Perugia and Universit\`a di Perugia,
     I-06100 Perugia, Italy \break
  \iPISA
     INFN Sezione di Pisa and Universit\`a di Pisa,
     I-56100 Pisa, Italy \break
  \iRUT
     Rutgers University,
     Piscataway, New Jersey 08855 \break
  \iRAL
     Rutherford Appleton Laboratory,
     Chilton, Didcot, Oxon OX11 0QX United Kingdom \break
  \iSOGANG
     Sogang University,
     Seoul, Korea \break
  \iSLAC
     Stanford Linear Accelerator Center, Stanford University,
     Stanford, California 94309 \break
  \iTENN
     University of Tennessee,
     Knoxville, Tennessee 37996 \break
  \iTOH
     Tohoku University,
     Sendai 980 Japan \break
  \iVAND
     Vanderbilt University,
     Nashville, Tennessee 37235 \break
  \iWASH
     University of Washington,
     Seattle, Washington 98195 \break
  \iWISC
     University of Wisconsin,
     Madison, Wisconsin 53706 \break
  \iYALE
     Yale University,
     New Haven, Connecticut 06511 \break
  \dead
     Deceased \break
  \andgen
     Also at the Universit\`a di Genova \break
  \andper
     Also at the Universit\`a di Perugia \break
\rm
%
 
\end{center}
 
\vspace{2.5cm}
 
\normalsize
 
\centerline{\bf ABSTRACT}
 
\noindent
We present the first study of rapidity gaps in \epem annihilations,
using \z0 decays collected by the SLD experiment at SLAC.
Our measured rapidity gap spectra fall exponentially
with increasing gap size over five decades, and we observe no anomalous
class of events containing large gaps. This
supports the interpretation of the large-gap events measured in
\pp and ep collisions in terms of exchange of color-singlet objects.
The presence of heavy
flavors or additional jets does not affect these conclusions.
 
\vfill
\eject
 

Since the initial observation of hadronic jets rapidity has been used
to characterize the momentum of particles in jets in a frame-invariant
manner.
The rapidity distribution has been studied in \epem
annihilation, ep and hadron-hadron collisions, and fixed-target
experiments, and is a characteristic of strong
interactions that is well described by perturbative QCD combined with
iterative models of jet fragmentation \cite{FF}.
Hard scattering of quarks or gluons can be modeled by
color fields between the outgoing partons that fragment
into the observed final-state particles,
typically populating the whole rapidity range.
 
Recently, exchange of color-singlet objects in hard
diffractive hadron-hadron scattering processes
characterized by events containing large gaps in the particle
rapidity spectrum has been discussed \cite{bj}.
Subsequent studies of \pp collisions at the Fermilab Tevatron collider
found that roughly 1\% of events comprising at least two
high-transverse-energy ($E_T$) jets contain a large rapidity region
between the two highest-$E_T$ jets with no particle activity
\cite{d01,cdf,d02}. These events have been interpreted in terms of
color-singlet exchange between the interacting partons.
Exchange of electroweak bosons is estimated to contribute only a small
fraction of the observed rate of gap events and a model incorporating
pomeron exchange is in agreement with the data \cite{cdf,d02}.
Large rapidity gaps have also been observed in
roughly 10\% of all photoproduced dijet
events in deep-inelastic scattering at the HERA ep collider
\cite{hera,zeus} and have also
been interpreted in terms of color-singlet exchange.
Models involving either vector meson dominance of the exchanged virtual
photon or hard diffractive scattering via pomerons
describe the data \cite{hera,zeus}.
 
Color exchange processes account successfully for the properties of the
majority of dijet events and may give rise to large rapidity gaps
due to random fluctuations. In both the ep and \pp
cases the interpretation of rapidity-gap events
in terms of color-singlet
exchange depends upon an understanding of this color-exchange background.
In the \pp experiments this was estimated
by extrapolation of fits to the particle multiplicity distribution
in rapidity intervals into the zero-particle, or rapidity gap,
region \cite{cdf,d02}. In the ep experiments the background was
estimated using both a Monte Carlo simulation and an \adhoc
parametrisation of color-exchange \cite{zeus}.
In both cases direct measurement of the spectrum of rapidity gaps arising
in color-exchange jet fragmentation is preferable and would
clarify the interpretation of the large-gap events.
 
Electron-positron annihilation into hadronic final states
provides an ideal laboratory for study of this issue as, in QCD,
it proceeds via creation of a primary quark and antiquark
connected by a color field which fragments into the observed hadrons.
The inclusive particle rapidity distribution has a broad plateau
centred at zero \cite{tasso}. Inclusive studies of the local
rapidity-density of particles have been performed
\cite{tassoint}, with the aim of investigating scale-invariant
cascade mechanisms in multiparticle production, but
no previous study has been made of the size of rapidity
gaps between adjacent particles. Large gaps are expected
to occur in \ep \ra \z0 \ra q$\overline{\rm q}$q$\overline{\rm q}$
or q$\overline{\rm q}$gg events when two color-singlet pairs of partons
are formed. The production rate of such events with a gap of at
least two units of rapidity is
estimated \cite{bjbrlu} to be about 10$^{-5}$ of all hadronic
\z0 decays, but with an unknown background from fluctuations
in \z0 \ra \qq fragmentation.
 
Here we present the first measurements of rapidity gaps
in \epem \ra hadrons, using \z0 decays.
We compare our results with the JETSET 7.4 \cite{jetset}
jet fragmentation model and with the perturbative QCD prediction
\cite{bjbrlu} for the rate of large-gap events.
We compare our results with
measurements from \pp collisions, where hadronic jets
are initiated by scattering of u and d quarks or gluons,
and with measurements from ep collisions, where jets are
initiated predominantly by u or d quarks.
We have isolated event samples enriched in
primary light (u,d,s) and heavy (b) quarks, and have measured rapidity
gap spectra in both. Finally, we
have studied the dependence of the rapidity gap spectra on
the event jet topology.

The \epem annihilation events produced at the $Z^0$ resonance
by the SLAC Linear Collider (SLC)
have been recorded using the SLC Large Detector (SLD) \cite{sld}.
This analysis used the charged tracks measured
in the central drift chamber \cite{cdc}
and in the vertex detector \cite{vxd}.
The trigger and initial hadronic event selection criteria are
described elsewhere \cite{sldalp}.
Events were required to have a minimum of 5 well-measured tracks
\cite{sldalp}, a thrust axis \cite{thrust} direction
within $|\cos\theta_T| < 0.71$, and a charged energy
calculated from the selected tracks assuming the $\pi^{\pm}$
mass of at least 20~GeV.
From our 1993-95 data samples 101,676 events passed these cuts,
including a background of
$0.3\pm 0.1\%$ dominated by \z0 \ra \tautau events.
 
Particle rapidity $y = 0.5 \ln (E+p_{\parallel})/(E-p_{\parallel})$,
where E is the particle energy and $p_{\parallel}$ its momentum
component along the thrust axis of the event, was
calculated from the measured momentum assuming the $\pi^{\pm}$ mass.
The charged track rapidity spectrum is shown in Fig. 1(a).
We ordered  the ${N}$ tracks in each event by their rapidity,
which defined ${N - 1}$ rapidity gaps, $\Delta\y$,
between pairs of adjacent tracks.
For each event we considered the largest gap, \Dymax, the
average gap, \Dy, and the size of the central gap, $\Delta\y_{cent}$,
\ie that gap containing the mean rapidity of the tracks.
The \Dymax distribution is shown in Fig. 1(b);
as \Dymax increases it rises to a peak around \Dymax = 1,
and for 1 $\leq$ \Dymax $\leq$ 6 it
falls approximately exponentially; for \Dymax $\geq$ 6 a
`shoulder' is apparent, suggesting that the data sample contains
a class of events characterized by large rapidity gaps.
 
This feature was found to be due to the 0.3\% contamination
of \tautau events in the \qq sample. The  \y and \Dymax
distributions for a sample of JETSET 7.4 \cite{jetset} \qq
and KORALZ 3.8 \cite{koralz} \tautau Monte Carlo events, combined
according to their Standard Model fractions of \z0 decays,
and subjected to a simulation
of the detector and to the same selection cuts as the data, are also
shown in Figs. 1(a) and 1(b)  respectively; the simulation
describes the data well.
The requirement on the minimum
number of charged tracks per event $n_{ch}$ was
increased from 5 to 7, yielding 100,964 events with
a reduced contribution from \tautau events of 0.13\%.
The resulting \Dymax distributions for the
data and simulated events are shown in Fig. 1(b);
for \Dymax $\geq$ 1.0 the data
fall exponentially and are well modelled by the simulation.
Similar results were obtained for \Dy and \Dycent (not
shown). All further analysis was based upon the requirement
$n_{ch}$ $\geq$ 7.
 
Also shown in Fig. 1(b) is the \Dymax distribution
from JETSET \qq events at the generator level; all particles, except
neutrinos, with lifetimes larger than 3 x $10^{-10}$ seconds,
as well as $\pi^0$s, were considered
stable and were used in the analysis. Comparison of this
with the measured distribution indicates the
effects of the detector
and of the track and event selection criteria, in particular the
exclusion of neutral particles, on the \Dymax spectrum.
The generator-level spectrum
exhibits the same structure as the data, and it
is shifted to smaller values of \Dymax and has a steeper exponential
decrease due to the larger number of particles considered in each event.
 
We quantified our data sample in terms of
the fraction of events $f$(\yzero)
containing no charged particles in the interval \absy $\leq$ $y_0/2$,
shown in Fig. 2; it falls exponentially.
The JETSET \qq simulation falls below the data for \yzero $\geq 3$,
but the data are well described by the simulation including \tautau
events, demonstrating that even for events with $n_{ch}$ $\geq$ 7
the influence of \tautau event contamination is noticeable at large
gap values. In Fig.~2 we
compare $f(y_0)$ with the perturbative QCD prediction
\cite{bjbrlu} for the fraction of \z0 decays into
q$\overline{\rm q}$q$\overline{\rm q}$ and q$\overline{\rm q}$gg,
where in each case two color-singlet parton-pairs are produced
with rapidity separation \yzero.
The prediction, for \alpmzsq = $0.120\pm0.008$ \cite{sldalp},
has a similar exponential falloff as the data,
but is between two and three orders of magnitude smaller.
The effect of neglect of neutral particles (Fig.~1)
is much smaller than this difference.
These results imply that the production of events
with large rapidity gaps in hadronic \z0 decays is
dominated by fluctuations in color-exchange jet fragmentation.
 
We also compare $f(y_0)$ with
similar measurements from \pp  and ep collisions, although,
due to differences in selected event topology, particle selection, and
definition of rapidity, such a comparison can be made
in qualitative terms only. The D0 Collaboration has studied
\cite{d01} the fraction of events that have no tagged particles
in pseudorapidity space
between the two highest-$E_T$ jets in their multijet
event data sample.
Jets were defined using a cone algorithm \cite{d01} and required to
have $E_T$ $>$ 30 GeV; \yzero was taken to be the difference in
pseudorapidity between the edges of the two jet cones, and
a tagged particle was defined to be a tower in
the D0 electromagnetic calorimeter with $E_T$ $>$ 200 MeV.
For \yzero $\leq 1$ the D0 results decrease exponentially (Fig.~2);
for \yzero $>1$ they indicate a plateau,
with $f(y_0>3)$ = $0.0053\pm0.0009$ \cite{d01}.
The CDF Collaboration has reported
similar results, $f(0.8\leq y_0\leq4)$ =
$0.0085^{+0.0026}_{-0.0017}$ \cite{cdf}, based upon absence of
charged tracks of momentum larger than 400 MeV/$c$ between jets.
The ZEUS Collaboration has studied \cite{zeus}
photoproduced events containing two jets with $E_T$ $>$ 6 GeV.
The fraction of events that have no calorimeter clusters
with $E_T$ $>$ 250 MeV between the two jets, defined using a similar
cone algorithm as D0 but with a larger cone radius,
decreases with $y_0$ (Fig.~2) but may
reach a plateau; $f(y_0=1.7)$ = $0.11^{+0.02}_{-0.03}$ \cite{zeus}.
 
For $y_0<1$
our measurement of an exponentially falling rate of gap events
is qualitatively similar to both the \pp and ep cases at low $y_0$.
For $y_0>1$ our measured exponential decrease,
which is well described by the JETSET
model of color exchange between the primary quark and antiquark,
contrasts with the onset of a plateau in the \pp and ep data samples,
and supports the interpretation of the
large-gap events in \pp and ep collisions
in terms of hard colorless exchange processes \cite{d01,cdf,d02}.
 
The JADE jet-finding algorithm \cite{jade} was used to define jets and
the rapidity and gap distributions were studied as a function of
the jet multiplicity of the events.
We considered 2-jet and $\geq$3-jet events defined for
values of the scaled jet-pair
invariant mass in the range $0.005\leq y_{cut}\leq 0.13$.
We show results for the
12,394 2-jet events at $y_{cut}$ = 0.005 and the
6,885 $\geq$3-jet events at $y_{cut}$ = 0.13.
The former are events with two narrow back-to-back jets,
while the latter events contain at least 3 well-separated jets
where the additional jet(s) are typically due to hard gluon radiation.
 
The 2-jet rapidity spectrum (Fig.~3(a))
peaks at $|y|$ $\sim$ 2.5 and the region $|y|<2.5$
is depleted but non-zero, reflecting the color field
between the q and $\overline{\rm q}$. The $\geq$3-jet rapidity
spectrum peaks at a lower value, which is
expected from the lower jet energies, the jet topology, and
higher event track multiplicity. Corresponding
effects are visible in the \Dymax distributions (Fig. 3(b));
the 2-jet events peak at a gap of 1.5 units, and there
is a long tail at high gap values;
the 3-jet events peak at a gap size of
about 0.7 units and there are few events with \Dymax $>$ 2.5.
Similar results were observed at other values of $y_{cut}$ and
for the \Dy and \Dycent distributions (not shown).
In all cases the high-gap tails demonstrate exponential falloff
and are well described by hadronic Monte Carlo samples
subjected to the same selection criteria.
 
The rapidity and gap spectra were studied as a function of event
primary flavor using
impact parameters of charged tracks measured in the vertex detector
to tag samples enriched in light
($Z^0 \rightarrow u\bar{u}$, $d\bar{d}$ or $s\bar{s}$) and heavy
($Z^0 \rightarrow b\bar{b}$) flavor.
The 65,243 events containing no track with normalized
transverse impact parameter with respect to the interaction point
$b/\sigma_b>3$ were assigned to the light-flavor sample.
The 13,269 events containing three or more tracks with $b/\sigma_b>3$
were assigned to the heavy-flavor sample.
The light-flavor content of the light sample was estimated
to be 85\% and the b flavor content of the heavy sample
was estimated to be 89\%. Details of flavor tagging
are discussed in Ref.~\cite{mikeh}.
 
The rapidity spectrum of the b-tagged sample (Fig.~3(c)) is
relatively flat out to a $|y|\sim2.4$ and falls sharply thereafter.
The light quark
spectrum peaks at $|y|\sim0.6$ and falls more slowly at high
rapidity.
This difference can be explained by the fact that the rapidity of
B-hadrons in $Z^0$ decays peaks strongly at $|y| \approx 2.3$,
corresponding to the average B momentum of 32 GeV/$c$ \cite{brap},
and the large number of B-hadron decay products contributes
only a small additional spread about this value;
additional tracks from fragmentation following the B-hadron formation
are restricted to lower momenta and rapidities.
For \Dymax (Fig.~3(d)), as well as \Dy and \Dycent
(not shown), the distributions for the two flavor-tagged
samples are found to peak at roughly the same value, but the b-tagged
distribution falls faster
at high values.  These features are expected since the rapidity of
B-hadrons in \z0 decays is limited to 2.8, and their high decay
multiplicity populates the region $|y|<2.8$
independent of the fragmentation process.
The behavior of $f(y_0)$ is similar
for both flavor-tagged samples (not shown),
although the slope of the exponential
decrease is greater in the b-tagged case.
 
In conclusion,
we have studied distributions of rapidity and of rapidity gaps
using charged particles in \epem \ra \z0 \ra \qq decays, and their
dependence on the jet topology and quark flavor of the events.
After accounting for backgrounds from \tautau events,
we find that the inclusive distribution of the maximum rapidity
gap per event peaks at about 1 unit and subsequently
falls exponentially with increasing gap size.
In contrast to results from \pp and ep collisions
we find no evidence for any anomalous class of events characterized by
large rapidity gaps, and thus
support the interpretation of the large-gap \pp and ep events in
terms of exchange of color-singlet objects.
Differences in peak position and exponent of the falloff between samples
of different jet multiplicity
or flavor can be explained by differences in event topology, charged
multiplicity, and the effects of B-hadron production and decay.
 
We thank the personnel of the SLAC accelerator department
and the technical staffs of our collaborating institutions for
the successful operation of the SLC and
the SLD. We thank J.D.~Bjorken, S.J.~Brodsky and B.~May for helpful
discussions.
 
\vfill
\eject

\vskip 2truecm
 
\section*{Figure captions }
 
\noindent
{\bf Fig. 1:}
(a) Normalized distribution of the absolute rapidity of charged
particles;
data (solid circles) and Monte Carlo simulation (histogram).
(b) Normalized distribution of event maximum rapidity gap.
For $n_{ch}\geq5$ ($n_{ch}\geq7$)
the data are shown as open (solid) circles and the combined \qq and
\tautau Monte Carlo sample is shown as the dot-dashed (dashed)
histogram.
The solid histogram is the distribution for Monte Carlo \qq events at the
generator level.
 
\noindent
{\bf Fig. 2:}
Fraction of events containing no charged particle within
\absy $\leq$ $y_0/2$ (solid circles).
Prediction from JETSET \qq events (dashed histogram), and
from the combined sample of \qq and \tautau events (solid histogram).
The hatched region is the prediction from Ref.~\cite{bjbrlu}.
Similar results are shown from D0 \cite{d01} (squares)
and ZEUS \cite{zeus} (open circles).
 
\noindent
{\bf Fig. 3:}
Normalized distributions of (a) charged particle
rapidity and (b) event maximum rapidity gap for 2-jet
(solid circles) and $\geq$3-jet (open circles) events;
the simulations are shown as
solid (2-jet events) and dashed ($\geq3$-jet events) histograms.
Normalized distributions of (c) charged particle rapidity and
(d) event maximum rapidity gap for u,d,s-- (solid circles)
and b--tagged (open circles) samples;
the simulations are shown as
solid (u,d,s--tagged) and dashed (b--tagged) histograms.
 
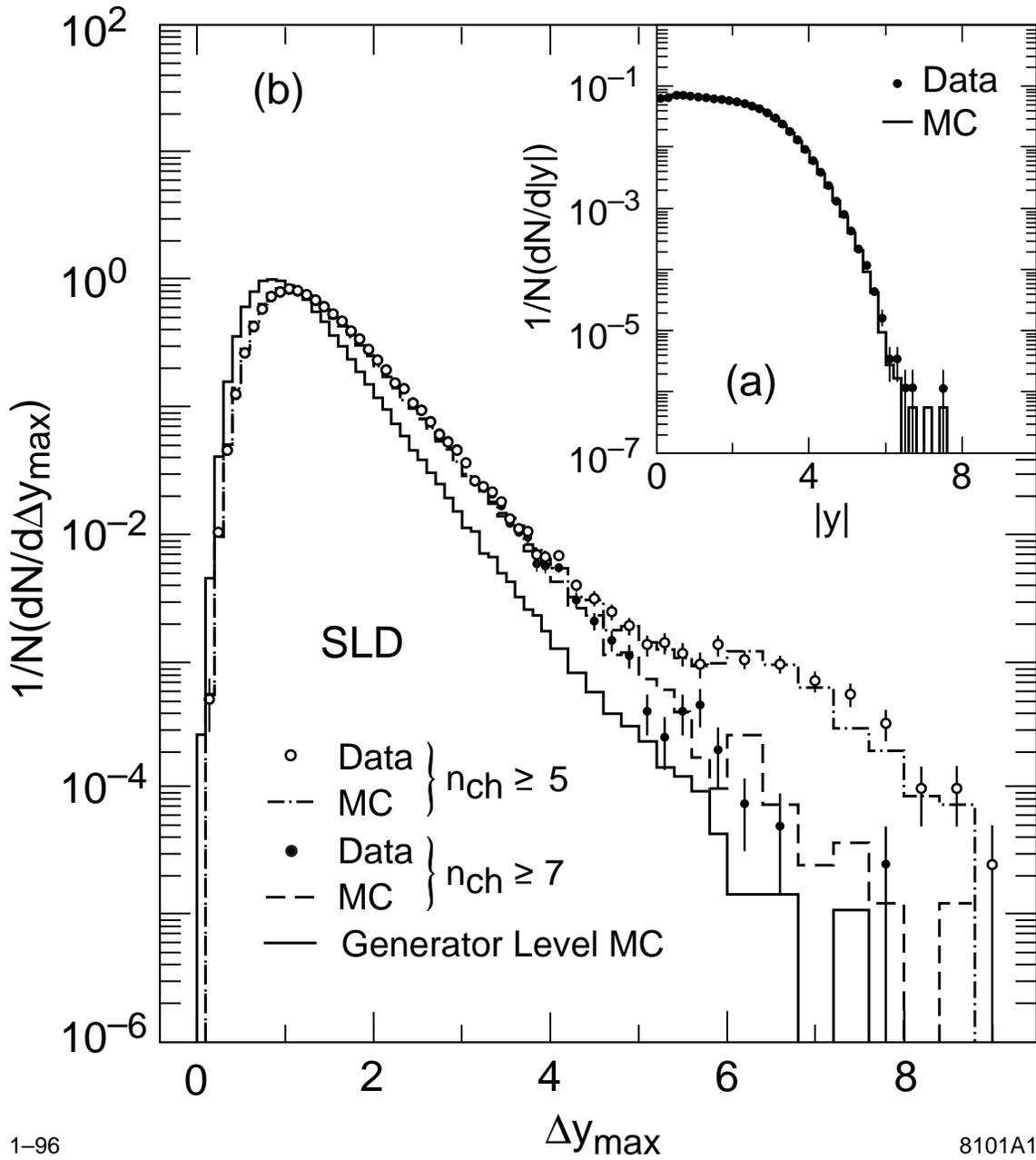
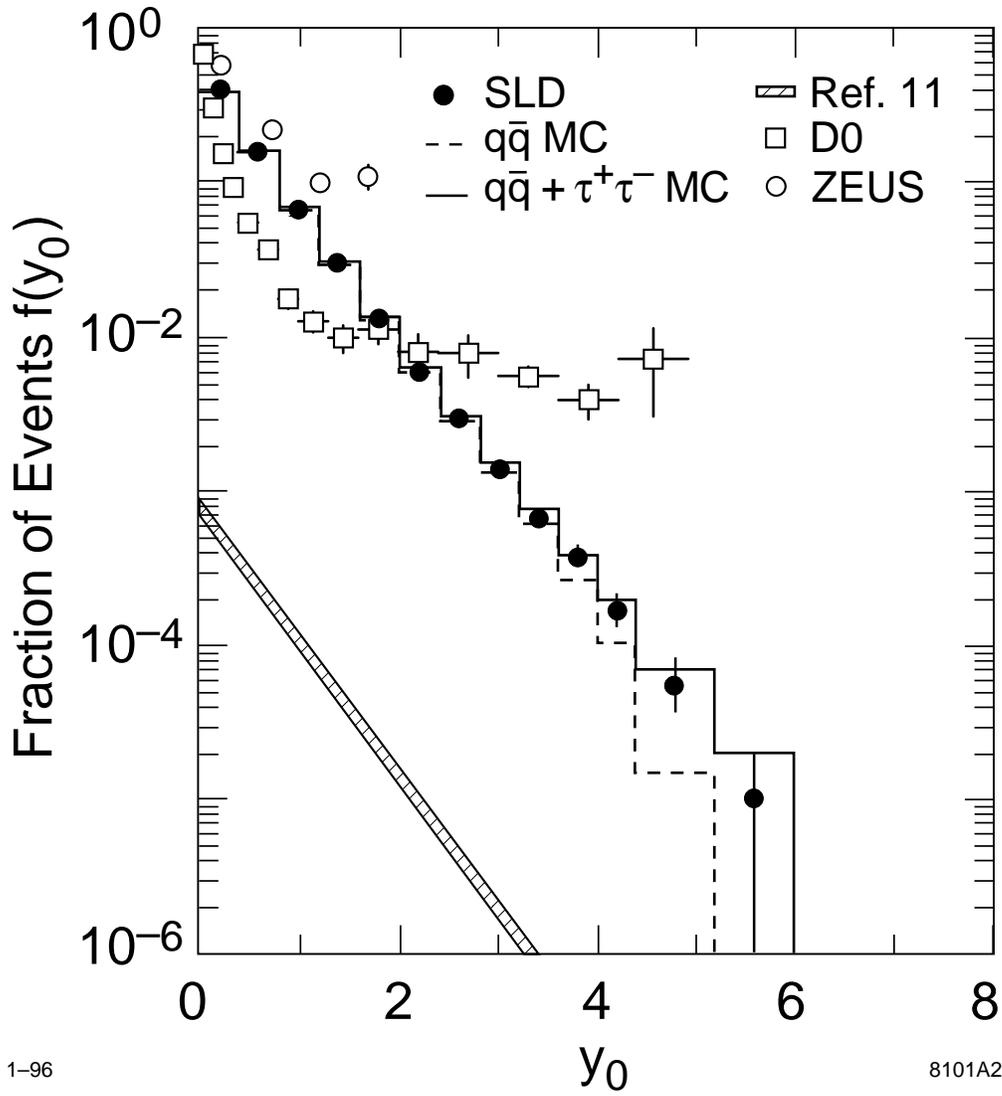
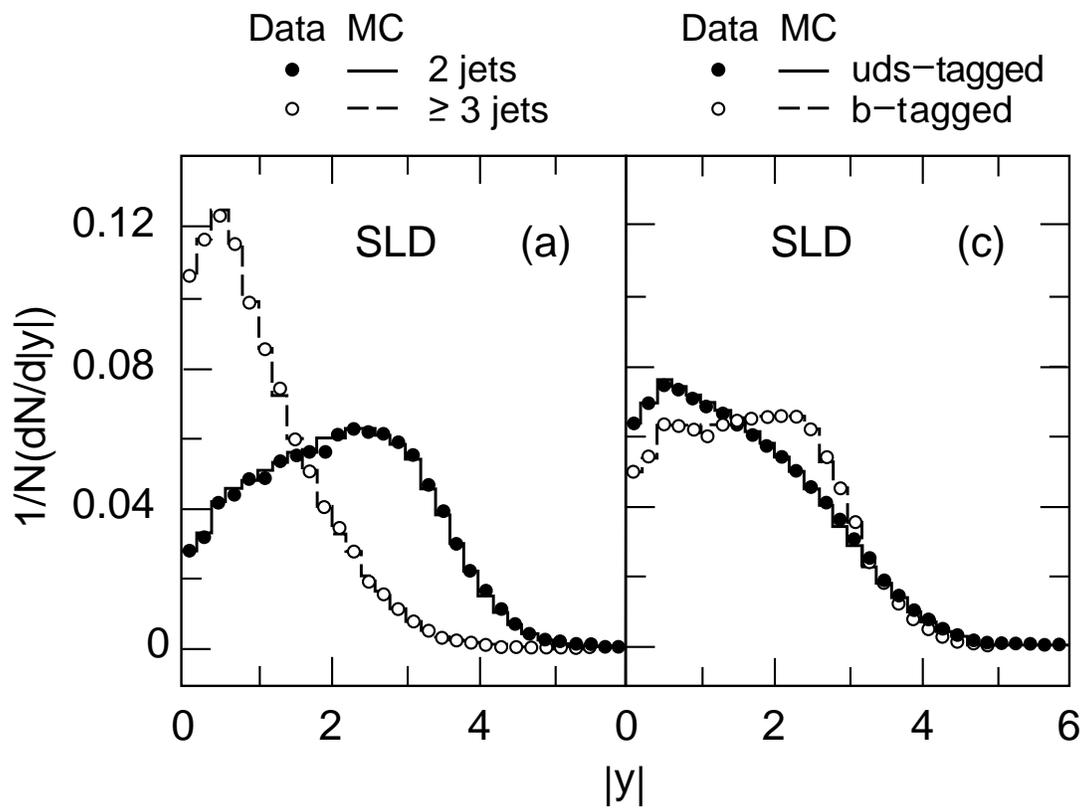
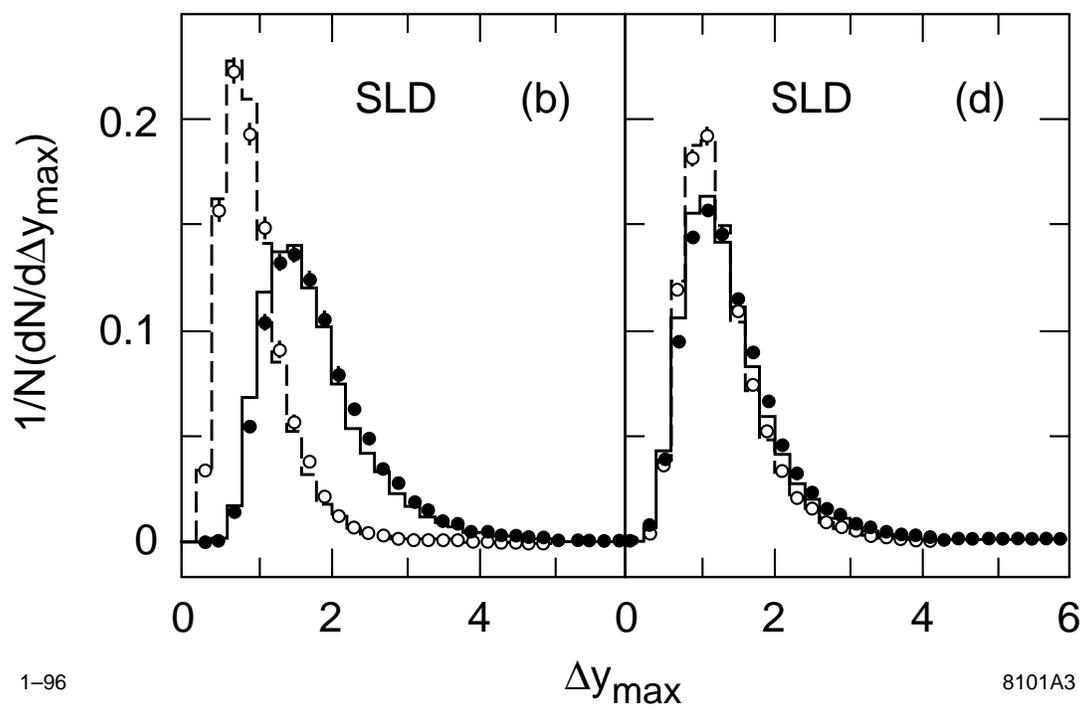
\end{document}